# Spatiotemporal Magnonic Vortex Beams with Alternating Transverse Orbital Angular Momentum

Muyang Xie(谢沐阳)[1*], Chenchen Liu(刘晨晨)[1*], Jian Huang(黄剑)[1], Zhenyu Wang(王振宇)[2], Xinwei Dong(董新伟)[1], and Ruifang Wang(王瑞方)[1**]

[1]Department of Physics, Xiamen University, Xiamen 361005, China

[2]School of physics and electronics, Hunan University, Changsha 410082, China

*authors contributed equally to this work

**corresponding author. Email: wangrf@xmu.edu.cn

Recent theoretical and experimental advances have demonstrated spatiotemporal photonic and acoustic vortex beams in free space. Such spatiotemporal vortex beams possess orbital angular momentum oriented perpendicular to the wave propagation direction. Herein, we report the discovery of spatiotemporal magnonic vortex beams in a confined ferromagnetic nanostrip geometry. The spatiotemporal magnonic vortex beam features stationary phase dislocations and exhibits wave propagation along a zigzag-like trajectory. Notably, the transverse orbital angular momentum carried by these phase dislocations displays spatial alternation. Our results differ distinctly from their photonic and acoustic counterparts, offering new insights into the fundamental research of spatiotemporal vortex beams.

**Keywords:** spatiotemporal magnonic vortex beams, phase singularity, transverse orbital angular momentum, spin waves

**PACS:** 75.78.-n, 75.78.cd, 75.40.Gb, 75.75.Jn

Vortex beams with orbital angular momentum (OAM) have been extensively investigated in photonics [1-5], acoustics[6,7], electronics[8-11] and magnonics[12-14]. The orientation of OAM is intrinsically linked to the spatial structure of the phase front. Conventionally, vortex beams with longitudinal OAM exhibit a characteristic screw phase dislocation[15,16] aligned along the propagation axis, with vanishing wave amplitude at the beam core. Their phase profile varies with the azimuthal angle $\varphi$ in the form of $e^{il\varphi}$, where the integer $l$ denotes the topological charge of the vortex. Longitudinal OAM is collinear with the beam axis and takes a quantized value of $l\hbar$. Experimental and theoretical studies[3,6,7,17] have verified that OAM can drive microscopic particles into circular motion. Furthermore, the achievable large OAM values open up new technological avenues for information transmission[2,4,9,10].

In contrast to conventional vortex beams, which exist in 3D space or axisymmetric media, spatiotemporal (ST) vortex beams can propagate in 2D space, with their wave fields carrying transverse OAM[3,18,19] that is orthogonal to the propagation direction. In this work, we report ST magnonic vortex beams in a ferromagnetic nanostrip with a vortex domain wall at its center. When magnonic plane waves pass through the vortex domain wall, the spin-wave propagates along a "zigzag-like" trajectory. The tilting of wave fronts is nonuniform in both the spatial and temporal domains. Four phase singularities of edge, or a mixture of edge and screw, type are observed, all of which are stationary. Moreover, the transverse OAM oscillates spatially between $\hbar$ and $-\hbar$.

We consider a ferromagnetic nanostrip lying in the *x-y* plane, with its magnetization direction denoted by unit vector $\boldsymbol{m}$ as illustrated in Fig. 1(a). The magnetization dynamic of the nanostrip is governed by the Landau-Lifshitz-Gilbert (LLG) equation:

$$\frac{\partial \boldsymbol{m}}{\partial t} = -\gamma \boldsymbol{m} \times \mathbf{H}_{eff} + \alpha \boldsymbol{m} \times \frac{\partial \boldsymbol{m}}{\partial t}, \tag{1}$$

where $\gamma$ and $\alpha$ are the gyromagnetic ratio and the Gilbert damping, respectively.

The sample has a length of 2500 nm, a width of 202 nm and a thickness of 5 nm. To investigate the magnetization dynamics, micromagnetic simulations were performed using the OOMMF code[20]. Material parameters corresponding to permalloy ($Ni_{0.8}Fe_{0.2}$) were adopted: magnetocrystalline anisotropy constant $K$ = 0 J/m$^3$, exchange coupling constant $A = 1.3 \times 10^{-11}$ J/m, saturation magnetization $M_s = 8.0 \times 10^5$ A/m, and the damping constant $\alpha = 0.01$. In the simulations, the sample was discretized with a cell size of $2 \times 2 \times 5\, nm^3$. In the initial magnetization state, a stable tail-to-tail vortex domain wall is located at the center of the nanostrip. On the left (or right) side of the wall, the magnetization is nearly uniform and aligned along the $-\hat{\boldsymbol{e}}_x$ (or $+\hat{\boldsymbol{e}}_x$) direction. Inside the vortex domain wall, the magnetization curls clockwise (circularity $c = -1$) in the x-y plane, and turns to the $+\hat{\boldsymbol{e}}_z$ direction at the core (polarity $p = +1$), as illustrated in Figs 1(a) and 1(b). To suppress spin-wave reflection, the damping constant $\alpha$ gradually increases from 0.02 to 1.0 at the left and right ends of the nanostrip. The triangular shape of the ends is designed to maintain magnetization uniform in these regions.

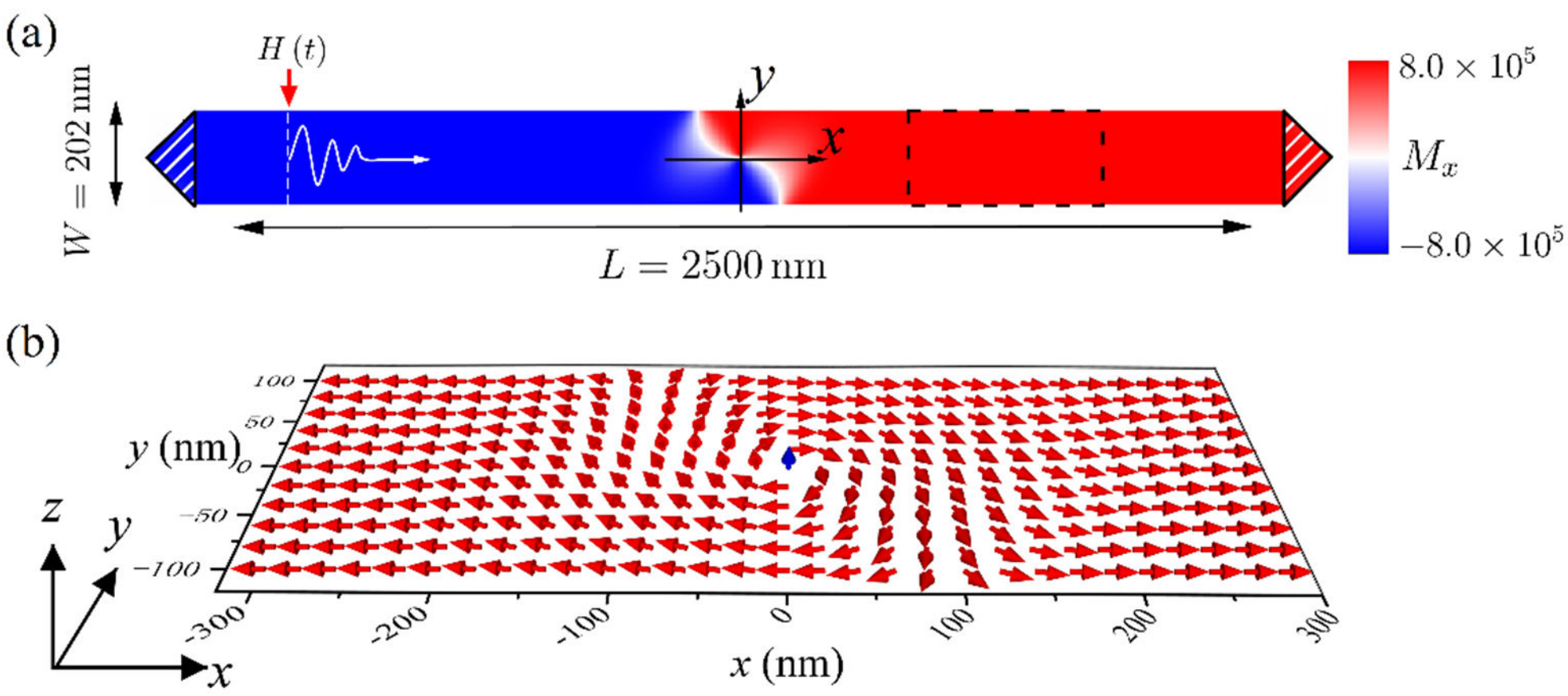

**Fig. 1.** Illustration of the initial magnetization state of the nanostrip. (a) A tail-to-tail vortex domain wall is located at the center of the sample. The left and right ends of the nanostrip, highlighted as black triangles with white strips, are designed to gradually shrink in width to reduce the amount of magnetic charges and maintain uniform magnetization. To minimize spin-wave reflection, higher damping constants varying from 0.02 to 1.0 are employed in the left and right ends. The black dashed box denotes the region where fast Fourier transform (FFT) calculations are performed. (b) Snapshot of the magnetization configuration of the vortex domain wall. The red arrows depict the magnetization direction. The vortex wall exhibits clockwise magnetization circulation and upward polarity at its core.

The dynamical magnetization $\boldsymbol{m}$ can be decomposed as $\boldsymbol{m} = \boldsymbol{m}_0 + \Delta\boldsymbol{m}$, where $\boldsymbol{m}_0$ is the static magnetization and $\Delta\boldsymbol{m}$ ($\ll \boldsymbol{m}_0$) represents the spin-wave excitation part[21,22]. Except for the region of the vortex domain wall, the static magnetization $\boldsymbol{m}_0$ is aligned along $\pm\hat{\boldsymbol{e}}_x$, and the *x*-component of $\Delta\boldsymbol{m}$ is nearly zero. Thus, the spin-wave excitation can be approximated as $\Delta\boldsymbol{m} =$ $(0, \Delta m_y, \Delta m_z)$. Furthermore, the excitation field oriented along $\hat{\boldsymbol{e}}_y$ results in a much larger $\Delta m_y$ than $\Delta m_z$.

Previous studies have shown that after a planar spin wave passes through a $180^{\circ}$ magnetic domain wall, its phase undergoes a $90^{\circ}$ phase shift[23,24]. Side jump and skew scattering have been reported for spin-waves propagating across magnetic skyrmions and chiral domain walls[21,22]. To date, few studies have focused on the spin-wave modes generated after a magnonic plane wave passes through a magnetic texture with inhomogeneous magnetization.

To investigate the spin-wave modes in the nanostrip, a sinc-function excitation field with the form $\boldsymbol{B}_s = \hat{\boldsymbol{e}}_y B_s \sin[2\pi f(t-t_0)]/[2\pi f(t-t_0)]$ was applied to a narrow region ($-950\ \mathrm{nm} \leq x \leq -940\ \mathrm{nm}$) for 10 ns, where $B_s = 50\ mT$, $f = 50\ \mathrm{GHz}$ and $t_0 = 200\ ps$. Fast Fourier transformation (FFT) was performed on the average *y*-component of the spin-wave excitation $\langle \Delta m_y \rangle$ in the region $350\ \mathrm{nm} \leq x \leq 750\ \mathrm{nm}$, i.e., after the spin-waves propagated through the vortex domain wall, as illustrated in Fig. 1(a). As shown in Fig. 2, the FFT amplitude spectrum reveals spin-wave modes at frequencies of 4.7, 5.4, 6.6, 7.5, 9.2, 10.9, and 11.5 GHz. The spatial distributions of these modes are presented in Figs. S1 and S2 of the supplemental data. The first four modes correspond to conventional plane waves, whereas the mode at 9.2 GHz exhibits distinct twisted wavefronts that rotate in the x-y plane. The resonant modes at 10.9 GHz and 11.5 GHz exhibit similar overall characteristics but possess nodal points along the width of the nanostrip. Owing to their more complex spatial field distributions, the analysis of these two modes is considerably more intricate. Accordingly, the present work mainly focuses on the mode at 9.2 GHz.

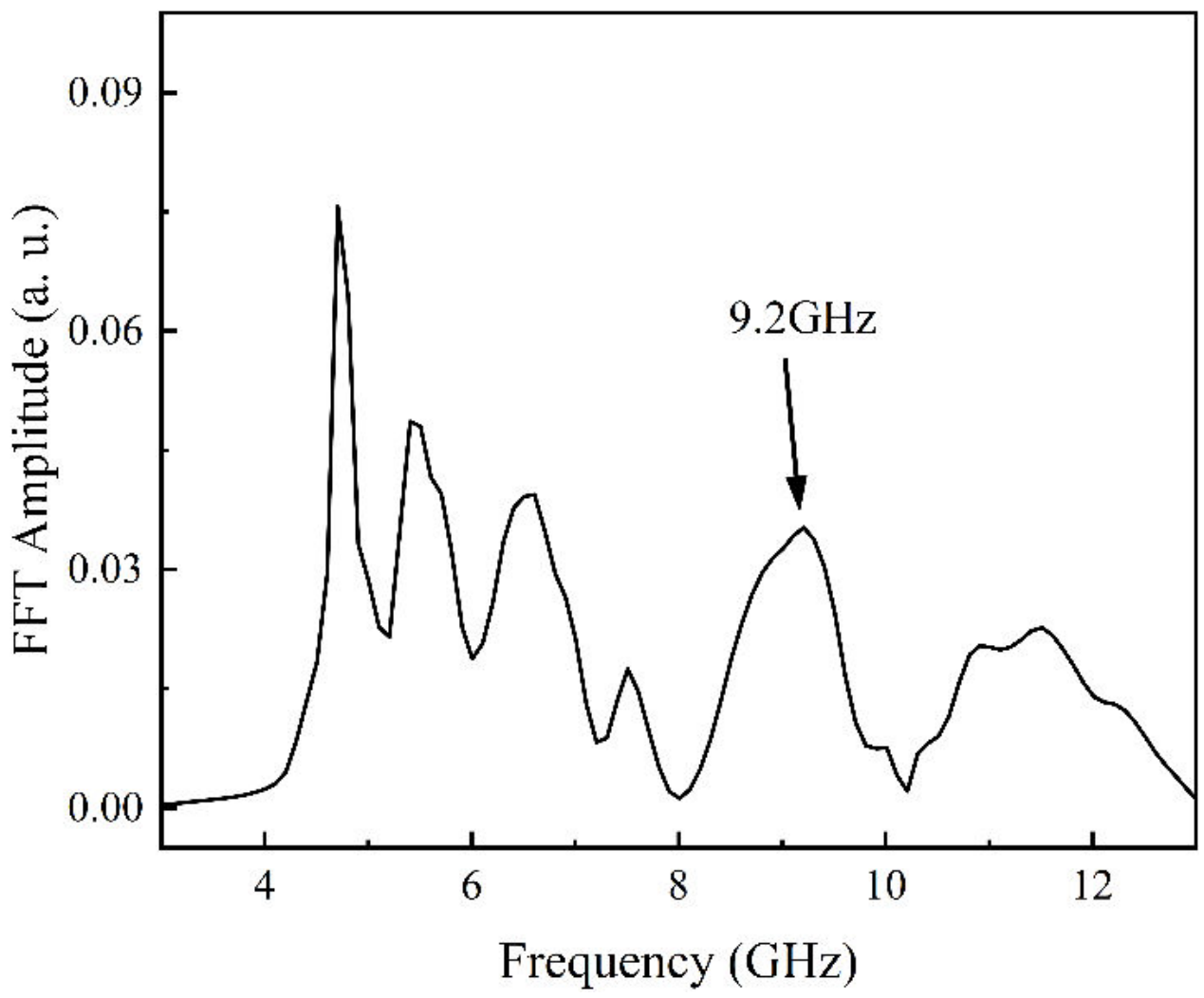

**Fig. 2.** FFT spectrum of spin-waves after passing through the vortex domain wall. The spectrum exhibits several distinct spin-wave modes, among which the 9.2 GHz mode is identified as a spatiotemporal magnonic vortex beam and is investigated in detail in this work.

To investigate the propagation of continuous spin waves, a sinusoidal field $\vec{B}(\mathrm{t}) = \hat{e}_y B sin2\pi f t$ (with $f = 9.2$ GHz and $B = 50\ mT$), is applied to the region $-950$ nm $\leq x \leq -940$ nm for a duration of 10 ns. This excitation field generates spin waves with parallel wavefronts (i.e., plane waves), which subsequently propagate until scattered by the vortex domain wall, as illustrated in Fig. 3. Local spherical coordinates $(\hat{\boldsymbol{e}}_r, \hat{\boldsymbol{e}}_\theta, \hat{\boldsymbol{e}}_\varphi)$ are defined with $\hat{\boldsymbol{e}}_r$ aligned to the static magnetization $\boldsymbol{m}_0$. The spin-wave excitation can be decomposed as[21] $\Delta\boldsymbol{m} = \Delta m_\theta \hat{\boldsymbol{e}}_\theta + \Delta m_\varphi \hat{\boldsymbol{e}}_\varphi$, and further represented by a complex wave function $\psi(\boldsymbol{r}, t) = \Delta m_\theta - i\Delta m_\varphi$. The spin-wave flux[21] is then defined as $\boldsymbol{j} = -iA(\psi^* \nabla\psi - \psi\nabla\psi^*)$. Before interacting with the vortex domain wall, the magnonic plane wave propagates linearly along the $\hat{\boldsymbol{e}}_x$ direction. The flux intensity $|\boldsymbol{j}|$ exhibits a symmetric distribution across the width of the nanostrip, with its maximum magnitude located at the central position ($y = 0$). The spatially inhomogeneous magnetization configuration of the vortex domain wall induces splitting of the spin-wave flux, whereby considerably more flux is scattered toward the upper half of the nanostrip than the lower half. Flux singularities emerge at the vortex core and multiple adjacent positions around the domain wall, as depicted in Fig. 3.

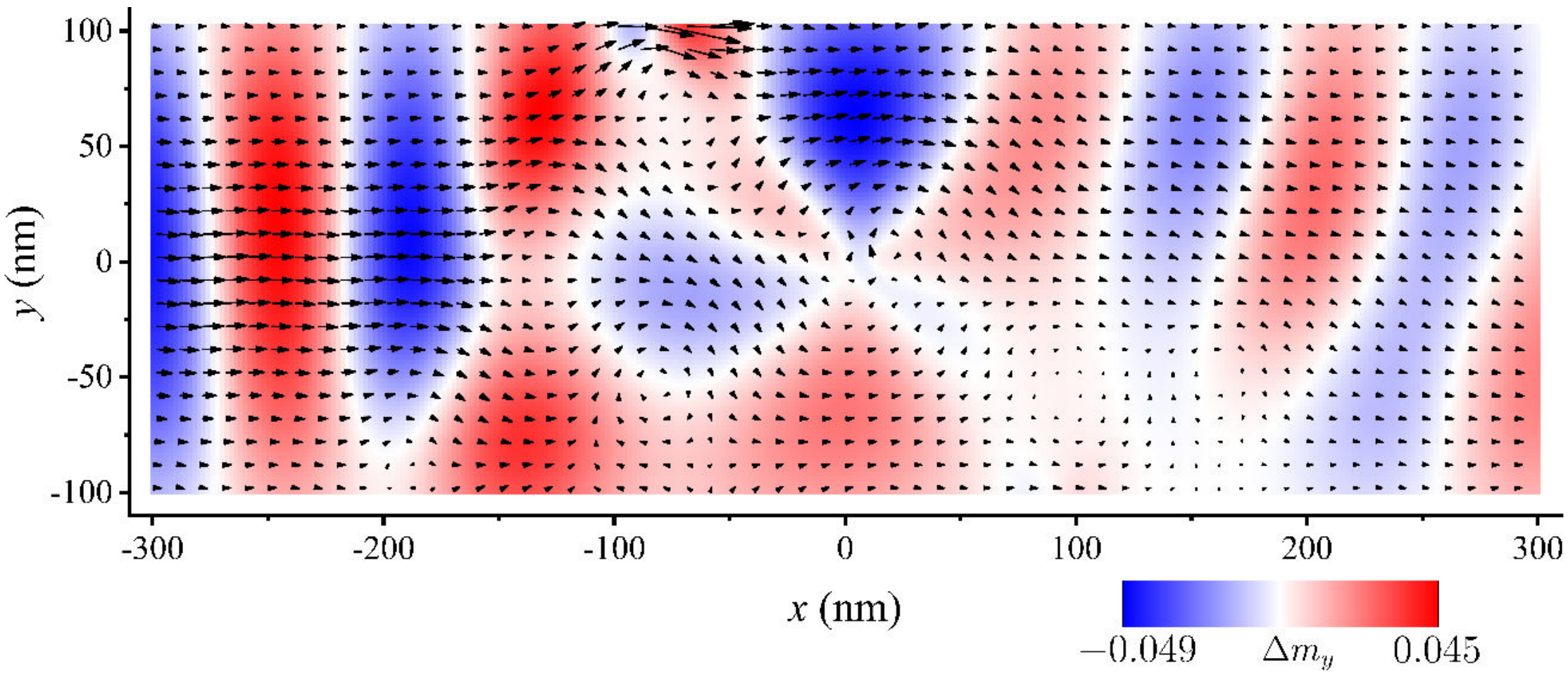

**Fig. 3.** Spin-wave scattering by the vortex domain wall. The background color indicates the spatial variation of $\Delta m_y$. The black arrows correspond to the local spin-wave flux $\boldsymbol{j}$, where the length of the arrows is proportional to the flux intensity.

After passing through the domain wall region, the spin wave propagates along a zigzag-like trajectory, as presented in Fig. 4(a). This behavior differs significantly from both the spiral propagation of vortex beams and the straight-line travel of ST optical beams[19,25,26], and acoustic beams[27] in free space. Meanwhile, the spatial distribution of $\Delta m_y$ closely follows this zigzag flux path. The spin-wave flux reaches its maximum at positions where the spin-wave excitation amplitude is the strongest. And distinct flux vortices can also be identified, with flux singularities and vanishing spin-wave excitation occurring at their cores. For the four flux vortices arranged from left to right in Fig. 4(a), their circularity $c$ varies regularly in the sequence $-1$, $+1$, $-1$ and $+1$, respectively. Such topological characteristics of these vortices are well consistent with the spin-wave flux evolution.

Figure 4(b) reveals the periodical tilting of phase fronts, which constitutes a characteristic topological feature of OAM-carrying waves. While the spin wave propagates within the *x*-*y* plane, its phase fronts rotate about the z-axis. Such rotation exhibits both spatial and temporal nonuniformity, giving rise to edge-type phase dislocations as well as mixed edge- and screw-type phase dislocations, as illustrated in Fig. 4(b).

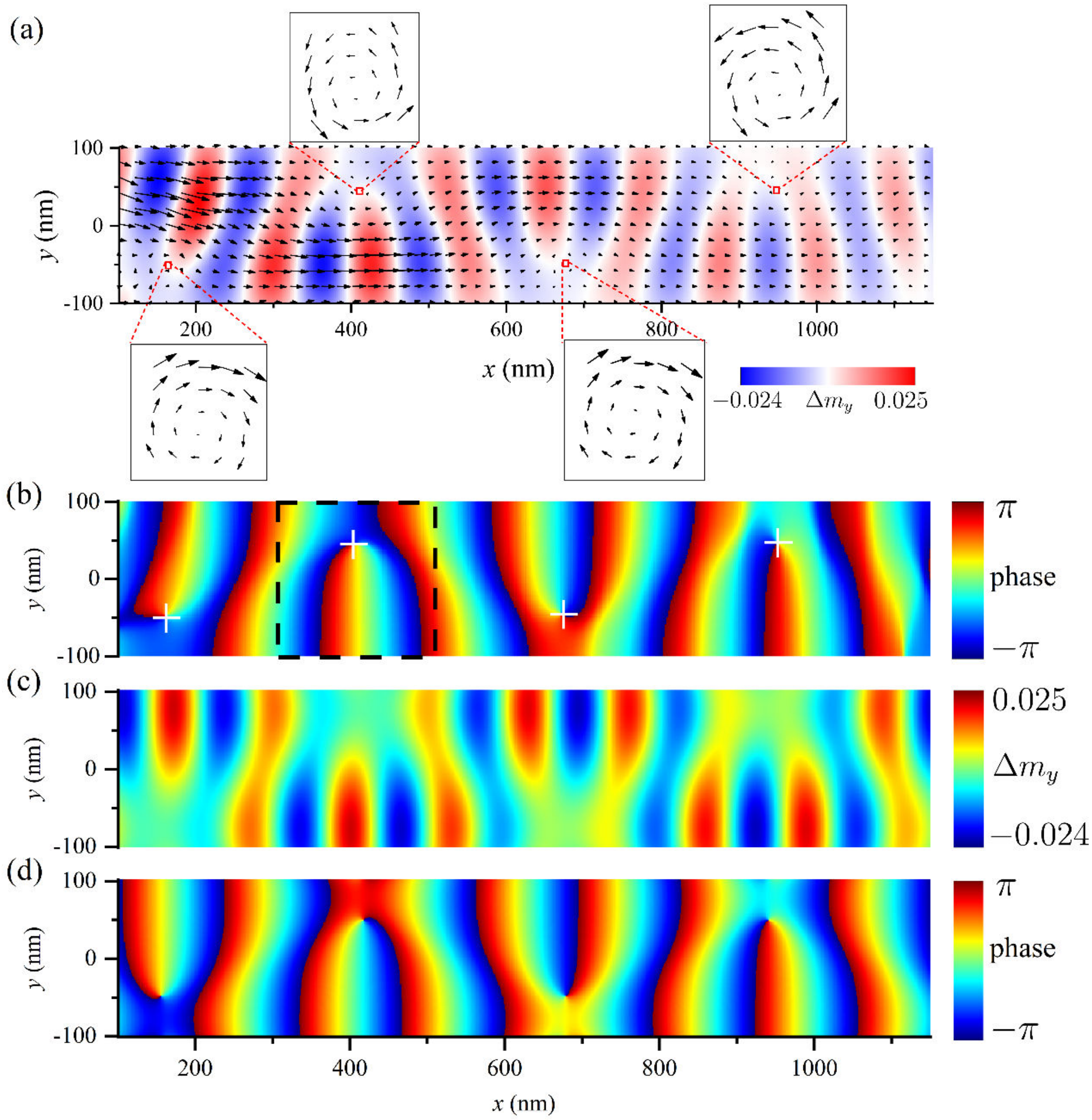

**Fig. 4.** Snapshots of the spin wave after its propagation through the vortex domain wall. The plots correspond to the instant at $t = 5000\ ps$, where $t = 0$ denotes the onset of the excitation field. In (a), the background color represents the micromagnetic simulation result of spatial variation of $\Delta m_y$. The spin-wave flux $\boldsymbol{j}$ is denoted by black arrows. The distribution of spin-wave excitation $\Delta m_y$ is in accordance with the zigzag-like path of spin-wave flux. (b) presents the micromagnetic simulation results of the phase distribution of the ST spin-wave excitation $\Delta m_y$. The tilted phase fronts give rise to four phase singularities, which are marked with white crosses; these phase singularities are either of edge type or a mixture of edge and screw types. Theoretical results of the $\Delta m_y$ distribution and the corresponding phase, calculated using Eq. (2), are displayed in (c) and (d), respectively.

To elucidate the generation of ST vortex pulses in free space, Bliokh et al. [18,28,30] constructed Bessel-type solutions of scalar waves via superposition of plane waves whose wavevectors are distributed along a circular contour. Given that the frequency of a plane wave is dependent on its wavevector, the ST vortex pulses constructed via this approach are inherently polychromatic, with their phase dislocations propagating along with the beam. By contrast, the vector spin-wave in this work propagates within the confined geometry of a magnetic nanostrip, where the phase dislocations remain stationary. As indicated by the spatial FFT analysis in Fig. S3 of the Supplementary Material, the ST magnonic vortex beam is composed of three plane waves with discrete wavevectors $\boldsymbol{k}_1$, $\boldsymbol{k}_2$, and $\boldsymbol{k}_3$, corresponding to the three distinct spots observed in **k**-space. The dynamic evolution of $\Delta m_y$ is formulated as:

$$\Delta m_y(x,y,t) = Re[Ae^{i(k_{x1}x-\omega t)} + Be^{i\left(k_{x2}x+k_{y2}y+\frac{\pi}{2}-\omega t\right)} + Be^{i\left(k_{x3}x+k_{y3}y-\frac{\pi}{2}-\omega t\right)}], \tag{2}$$

where $k_{x1} = 5.4 \times 10^7 m^{-1}$, $k_{x2} = k_{x3} = 4.2 \times 10^7 m^{-1}$, $k_{y2} = -k_{y3} = -2.0 \times 10^7 m^{-1}$, $\omega = 2\pi \times$ 9.2 GHz, and the amplitude ratio $A\!:\!B = 1.66$ is determined from the intensity of the spots in Fig. S3. All plane wave components are assumed to possess the same frequency of 9.2 GHz; otherwise, stationary phase dislocations cannot be formed. Based on Eq. (2), the theoretical distributions of $\Delta m_y$ and the corresponding phase at $t = 5000$ ps are plotted in Figs. 4(c) and 4(d), respectively, which show good consistency with the simulated results in Figs. 4(a) and 4(b).

The OAM carried by the spin-wave can be calculated[17] as

$$\boldsymbol{L} = \frac{\hbar}{S}\iint(\nabla\phi \times \boldsymbol{r})dxdy, \tag{3}$$

where $\phi = \arg(\Delta m_y + i\Delta m_z)$, $\boldsymbol{r}$ is measured relative to the phase singularity and $S$ is a small area surrounding the singularity, as illustrated in Fig. 5. Since both $\nabla\phi$ and $\boldsymbol{r}$ lie in the *x*-*y* plane, $\boldsymbol{L}$ is perpendicular to the *x*-*y* plane, indicating that the OAM is transverse in nature. By substituting the spin-wave excitation components $\Delta m_y$ and $\Delta m_z$ into Eq. (3) and setting $S$ as a circle with a radius ranging from 10 nm to 50 nm, we consistently obtain $L = -\hbar,\ +\hbar,\ -\hbar,\ +\hbar$ for the four phase singularities from left to right, respectively. Notably, at each phase vortex, the sign of the transverse OAM is always identical to the circularity of the spin-wave flux. Interestingly, the OAM values vary periodically in both the *x* and *y* dimensions. The temporal evolution of one phase dislocation, marked by a dashed black box in Fig. 4(b), is presented in Fig. 6. A similar phenomenon, termed “temporal diffraction”, has been previously reported in ST photonic and acoustic waves [18,19,27,28]. It can be observed that the structure of the phase dislocation undergoes periodic changes with a period $T = 110$ ps, which corresponds to the period of a 9.2 GHz wave. Remarkably, all phase dislocations remain stationary while the spin wave propagates within the nanostrip. In contrast, phase dislocations in ST optical and acoustic beams move along with the wave [18,27,29]. Micromagnetic simulations further reveal that the OAM of ST magnonic vortex beams can be switched by tuning the circularity of the magnetic vortex domain wall from $c = -1$ to $c = +1$. Notably, the OAM remains unaffected by the vortex polarity, as demonstrated by data not shown here.

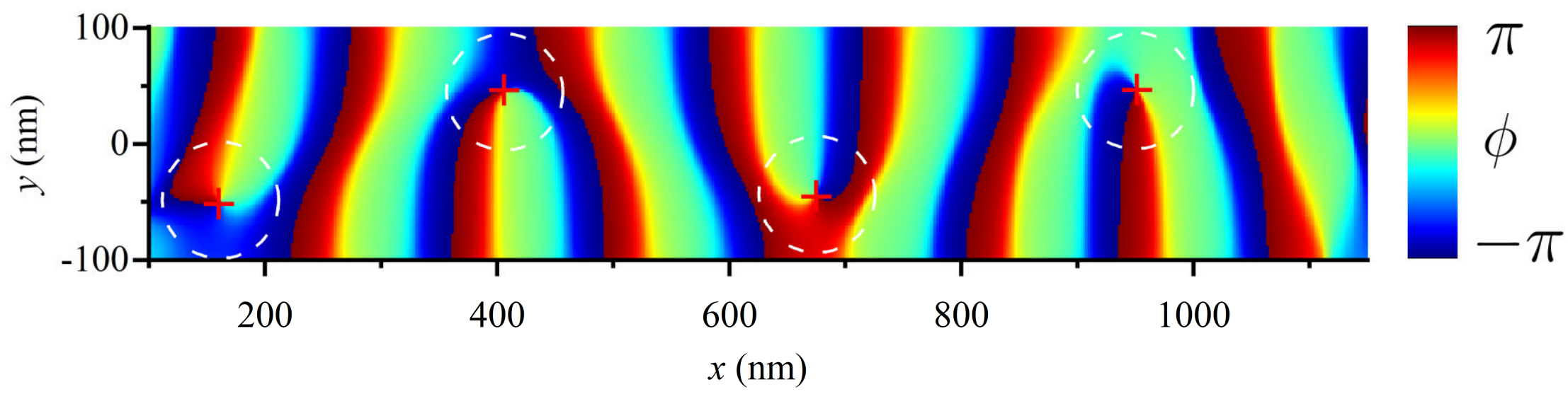

**Fig. 5.** The spatial distribution of $\phi = \arg(\Delta m_y + i\Delta m_z)$. Four phase vortices are marked with red crosses, and the white dashed circle denotes the integration area used to calculate the OAM via Equation (3).

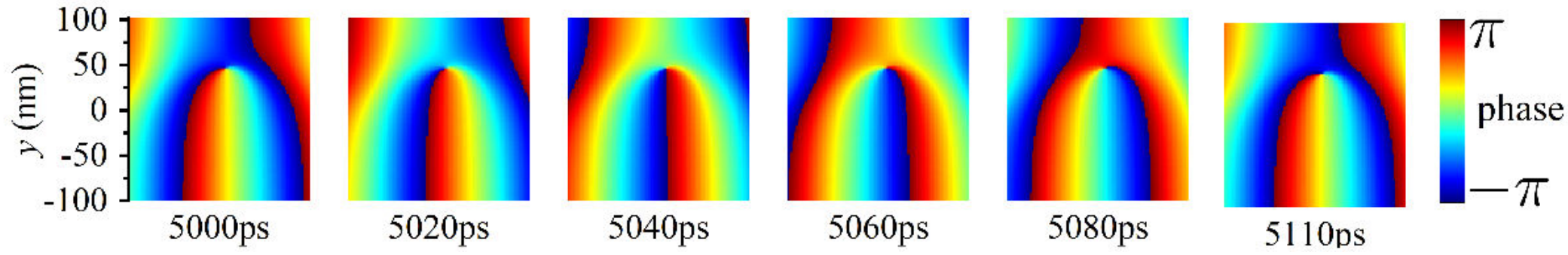

**Fig. 6.** Temporal evolution of the phase dislocation enclosed by the dashed black box in Fig. 4(b). The structural variation of the phase dislocation undergoes a complete cycle with a period of 110 ps.

In conclusion, we have revealed the existence of ST magnonic vortex beams in a ferromagnetic nanostrip by propagating planar spin waves through a vortex domain wall. The ST spin-wave flux is found to travel along an unconventional zigzag trajectory. The phase singularities remain stationary, irrespective of spin-wave propagation along the nanostrip. The transverse OAMs associated with these phase dislocations evolve periodically along both the $x$ and $y$ dimensions. The phase dislocation structure exhibits periodic temporal evolution, which clearly demonstrates the temporal diffraction behavior of ST magnonic vortex beams. Our findings differ distinctly from previous investigations on spatiotemporal photonic and acoustic vortex beams in free space. Beyond vortex domain walls, low-dimensional ferromagnetic systems host a wealth of nontrivial magnetic textures[31], including magnetic skyrmions, magnetic vortices, Bloch points, antivortices, and merons. This suggests a promising prospect for further exploring the emerging physics of spatiotemporal magnonic vortex waves.

We acknowledge financial support by the National Natural Science Foundation of China under Grant No. 11174238.

Supplemental Material

# Spatiotemporal Magnonic Vortex Beams with Alternating Transverse Orbital Angular Momentum

Muyang Xie(谢沐阳)[1*], Chenchen Liu(刘晨晨)[1*], Jian Huang(黄剑)[1], Zhenyu Wang(王振宇)[2], Xinwei Dong(董新伟)[1], and Ruifang Wang(王瑞方)[1**]

[1]Department of Physics, Xiamen University, Xiamen 361005, China

[2]School of physics and electronics, Hunan University, Changsha 410082, China

*authors contributed equally to this work

**corresponding author. Email: wangrf@xmu.edu.cn

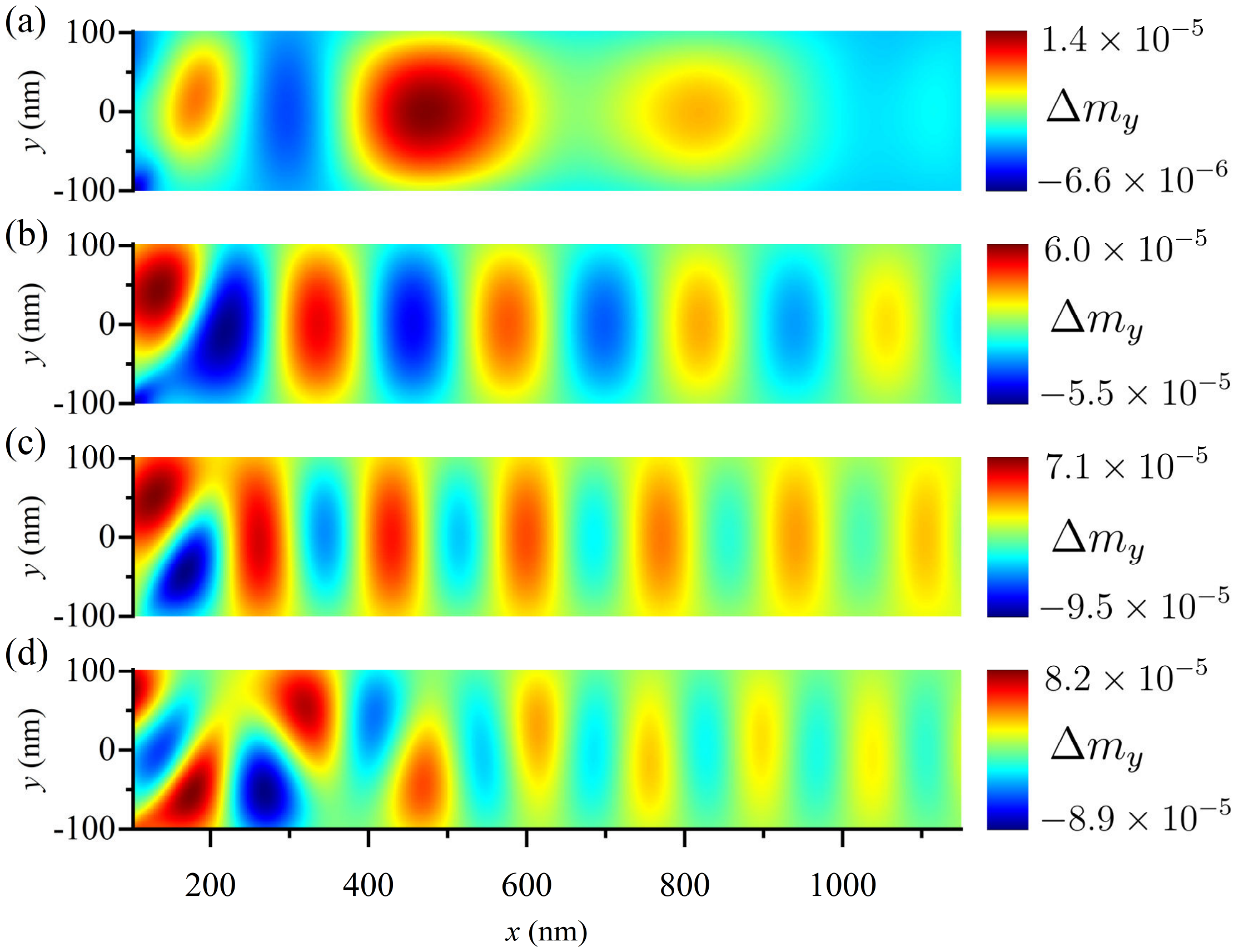

**Fig. S1.** Spatial variation of $\Delta m_y$ for the spin-wave modes at (a) 4.7 GHz, (b) 5.4 GHz, (c) 6.6 GHz, and (d) 7.5 GHz, respectively. Parallel wave fronts indicate that these modes are conventional plane waves.

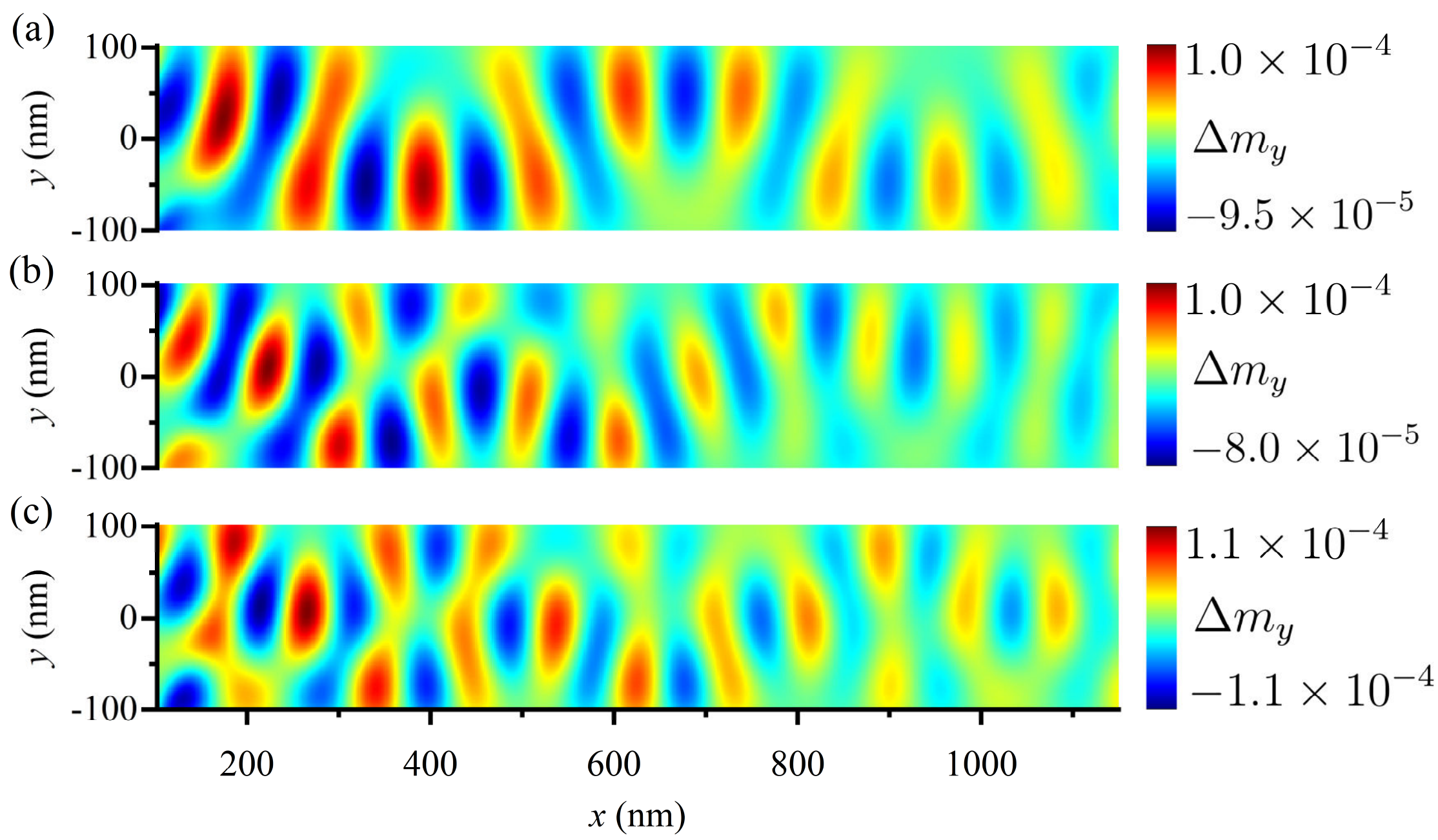

**Fig. S2.** Spatial variation of $\Delta m_y$ for the spin-wave modes at (a) 9.2 GHz, (b) 10.9 GHz, and (c) 11.5 GHz, respectively. The wave fronts tilt in the *x*-*y* plane, indicating that these modes carry transverse orbital angular momentum.

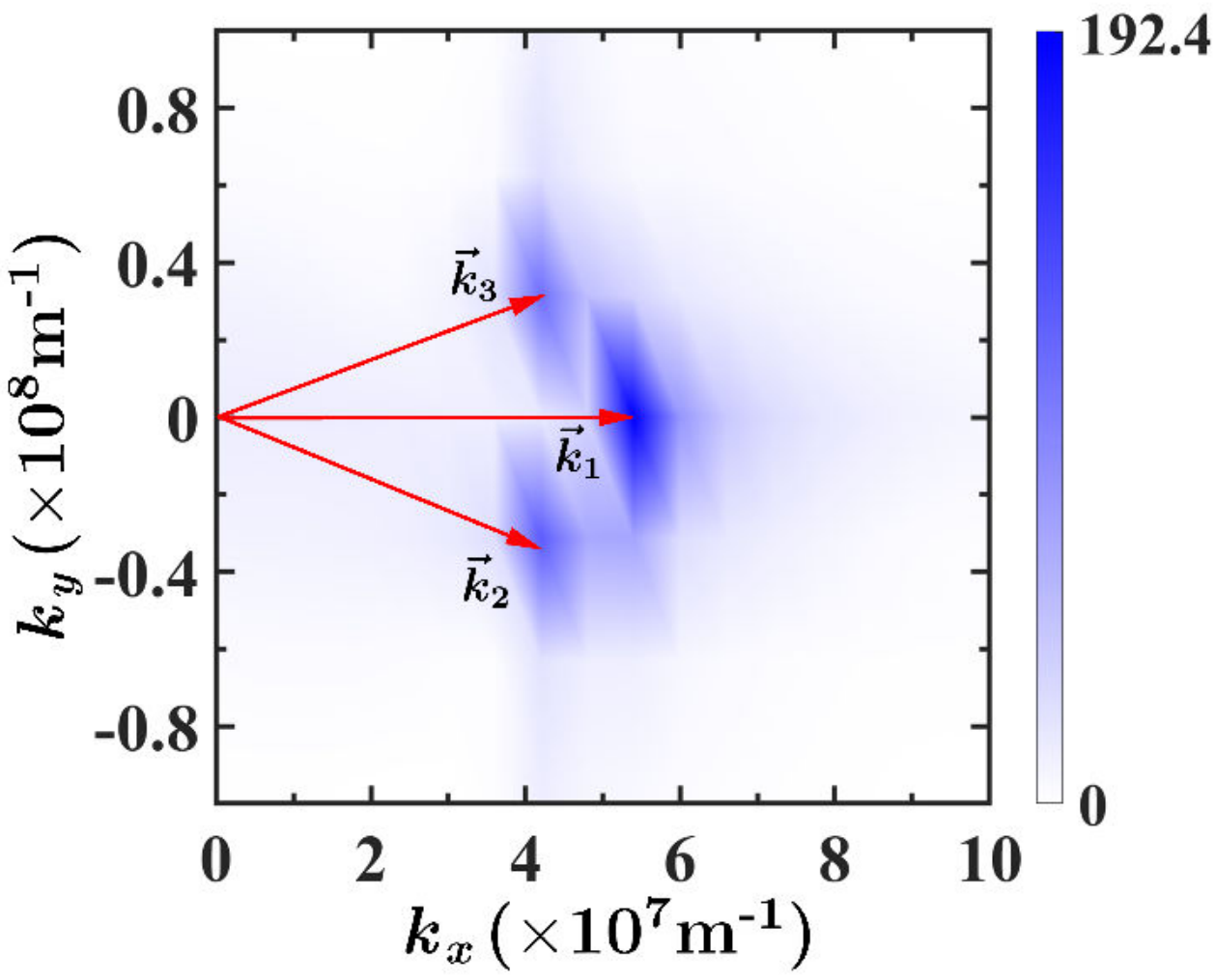

**Fig. S3.** Plane-wave spectrum of the ST magnonic vortex wave obtained by spatial FFT analysis. Intensity and wave vector of the three plane-wave components are used in equation (3) of the main text.